\documentclass[dvips]{article}

\usepackage{icrc2011}
\usepackage{color}
\usepackage{amsmath,amsfonts}
\def\Xmax{X_{\rm max}}

\title{Highlights from Telescope Array}

\shorttitle{Y.Tsunesada: Highlights from Telescope Array}

\authors{Yoshiki Tsunesada, for the Telescope Array Collaboration}
\afiliations{Tokyo Institute of Technology, Meguro, Tokyo 152-8550 Japan}
\email{tsunesada@cr.phys.titech.ac.jp}

\abstract{The results from the first three years observation in Telescope Array are reviewed.
The energy spectrum, mass composition, and anisotropy in arrival directions of ultra-high energy
cosmic rays are discussed. The energy spectrum of cosmic rays with energies above $10^{18}$ eV is determined
using the data of the fluorescence detectors, surface detectors, and with a hybrid shower reconstruction technique using
the both data. The ankle structure is clearly found at $10^{18.7}$ eV. The second bending point is also identified,
at which the flux of cosmic rays begins to decrease and the spectral index changes from $E^{-2.68}$ to $E^{-4.2}$.
The study of shower development using the fluorescence detector data shows that proton
is dominant in cosmic rays above $10^{18.2}$ eV. No apparent anisotropies in the arrival direction of cosmic rays
are found. I also present further experimental extensions which are being developed at the Telescope Array site.}


\begin{document}
\maketitle

\section{Introduction}
Telescope Array (TA) is a cosmic ray observatory of the largest exposure in the northern hemisphere,
located in the Millard County, 200 km southwest of Salt Lake City, Utah, USA. This is a hybrid experiment 
by collaborators from Japan, the US, Korea, Russia, and Belgium, 
using two types of detectors,
surface detectors (SDs) and fluorescence detectors (FDs). This can be regarded as a follow-up to
the AGASA and HiRes experiments, which used scintillation counters and fluorescence mirrors, to uncover the
mysteries of ultra-high energy cosmic rays and to clarify their origin.
In this paper, I review the major results from the TA first three years observation, including
the energy spectrum, mass composition, and anisotropy of ultra-high energy cosmic rays.
The activities for future extensions are also presented.

\section{TA detectors}
\begin{figure}
\begin{center}
\includegraphics[scale=0.43]{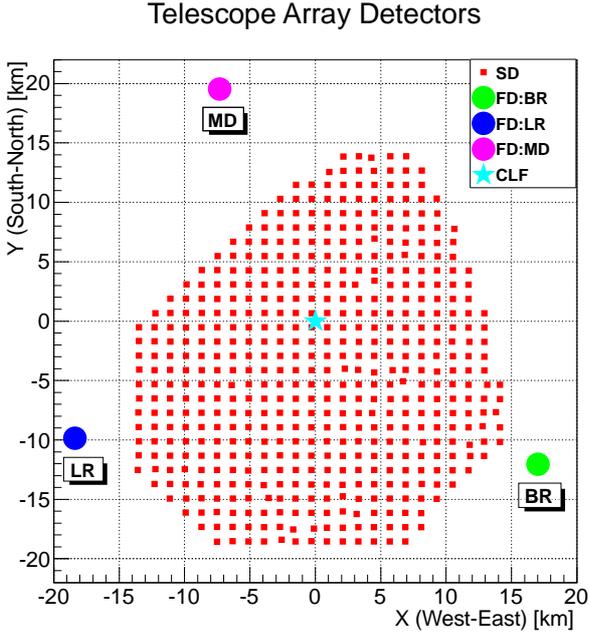}
\end{center}
\caption{Layout of the TA detectors.}
\label{fig:detectors}
\end{figure}

The TA-SD consists of 507 
three square meters double-layered scintillation counters in a square grid with 1.2 km separations 
in a total area of $\sim 700 {\rm km^2}$ (Figure \ref{fig:detectors}). The average atmospheric depth at the TA ground
level is $\sim 860 \, {\rm g/cm^2}$ ($\sim 1400$m above the sea level). Signals of shower particles
are sampled
at $50$ MHz frequency and digitized with 12-bit FADC, not integrating all charges from photo-tubes 
in a $\sim 10 \mu {\rm s}$ time constant as in conventional air shower array experiments like AGASA.
The SD data acquisition system is triggered by a coincidence,  within $8 \mu {\rm s}$
window, of neibouring three SDs with signals greater
than $3$ MIPs (minimum-ionizing particles) equivalent in their both layers. The SD array is fully efficient for cosmic rays with energies greater than $10^{18.8}$ eV \cite{IvanovSD}.

The fluorescence
detectors are installed in the three stations,  those are the Black Rock (BR, 12 telescopes), 
Long Ridge (LR, 12 telescopes), and Middle Drum (MD, 14 telescopes) sites,  viewing the sky above 
SDs. 
The 24 BR and LR telescopes were newly developed for TA, and the MD telescopes are 
refurbished HiRes-1 detectors. One BR/LR telescope is comprised of a cluster of photo-tubes (or a "camera"), 
and a reflecting mirror of $3.3$m diameter. The BR/LR camera has $16 \times 16$ 2-inch hexagonal tubes covering $15.5^{\circ} \times 18^{\circ}$ of the sky. A telescope triggering signal is generated
when five adjacent tubes in a camera are fired by incident photons, and waveform data of all the cameras are 
recorded in a station \cite{FDTameda}. The MD telescopes have $5.2 {\rm m^2}$ mirrors and cameras of
256 pixels using $40$mm Philips tubes. These telescopes are operated with the same electronics and data
acquisition system of the HiRes-1 experiment.
The total FD fields of view are 108, 108 and 114 degrees in azimuth
and $3 \sim 31$ degrees in elevation. See references \cite{NonakaSD,OgioFD} in this conference
for details of the TA detectors.

The TA detectors have been in operation since May 2008. The operational duty cycle of SD is 
$~\sim 98 \%$, and the total exposure is $\sim 2,700 \, {\rm [km^2 \, str \, yr]} (\sim 1.7 {\rm AGASA})$  ($\sim$ April 2011).
The telescope on-times are $\sim 2400$ hours (MD site),
which gives $\sim 1/3$ HiRes-1 exposure. The FD duty factor is $\sim 11\%$, and it reduces to $\sim 9\%$ after good weather cut.

\section{Shower analysis}
\subsection{Shower reconstruction with FD data}
The fluorescence detectors measure photons emitted from or around air showers with phototubes exposed
to the night sky to determine the longitudinal profiles.
The longitudinal profile of an air shower as a function of atmospheric depth $X$ is written by the Gaisser-Hillas function,
\begin{eqnarray}
 f(X) =  \left( \frac{X - X_0}{\Xmax - X_0} \right)^{\frac{\Xmax - X_0}{\lambda}} \exp \left( - \frac{X - \Xmax}{\lambda} \right) \label{eq:GH}
\end{eqnarray}
where $\Xmax$ is the depth at the maximum development of an air shower, $X_0$ is the first interaction point, $\lambda$ is a characteristic
length of particle interaction. This equation was originally proposed to describe the development of the number of charged particles \cite{GHpaper}, 
and is also applicable to an energy deposit profile which can be directly measured with FDs. 
The {\it calorimetric energy}
of the air shower is obtained by integrating the longitudinal profile, 
\begin{eqnarray}
 E_{\rm cal} =  N_{\rm max} \int  \left< \frac{{\rm d}E}{{\rm d}X} \right> f(X) {\rm d}X =  \int {\cal E}(X) {\rm d}X
\end{eqnarray}
where  $N_{\rm max}$ is the shower size at $\Xmax$, $\left< \frac{{\rm d}E}{{\rm d}X} \right>$ is the 
mean energy deposit of charged particles per unit length,
 and  ${\cal E}(X) \equiv {\cal E}_{\rm max} f(X)$ is the energy deposit profile of the shower. The primary energy of the cosmic ray which induced
 the shower is obtained from $E_{\rm cal}$ by taking into account the {\it missing energy} which is an
 {\it invisible} component of the shower energy carried away by neutral particles. 

The data analysis procedures for the MD data are almost identical to those used in the HiRes experiment \cite{DougMD}. For the
BR/LR data analysis, we have developed new codes suitable for the 10MHz sampling data acquisition system.

Directly calculating the number of charged particles using phototube signal and a given reconstructed air shower geometry is difficult due to the complexities of the FDs, as well as, accounting for the effects of Cherenkov light in the shower. For this reason, an 
{\it Inverse Monte Carlo} (IMC) method is employed. Using an IMC method, showers with varying $\Xmax$ and $N_{\rm max}$ (or ${\cal E}_{\rm max}$) are simulated and the number of measured photo-electrons is compared with that produced by the simulated showers. Simulation of the air shower allows for the treatment of Cherenkov radiation and ray-tracing methods are used to understand the acceptance of photons produced on the shower axis.
Here we can include
all the effects, such as photon shadowing by structures, mirror reflectances, and other calibration-related matters. 
We use {\it independently} developed FD simulation/reconstruction programs
utilizing the inverse Monte Carlo method. 
From the differences between energies determined by the
two codes, and together with $\sim5\%$ and $\sim3\%$ uncertainties 
which originate from non-predetermined primary nuclear types and the missing energy correction, 
the energy uncertainty from the reconstruction procedures is evaluated as $\sim 10\%$. Taking
into account the systematic uncertainties from FD calibrations \cite{TokunoCalib} 
or atmospheric effects \cite{TomidaLIDAR}, 
the total uncertainty in the TA FD energies is $\sim 22\%$ \cite{TsunesadaEnergyScale}.

%
%

\subsection{Measurement with the surface detector array}\label{sec:SDanalysis}
\begin{figure}
\begin{center}
\includegraphics[scale=0.4]{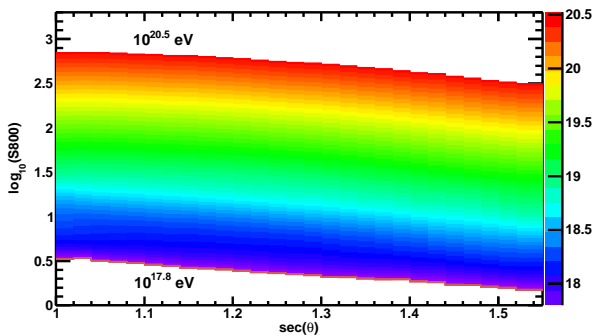}
\end{center}
\caption{The SD energy chart}
\label{fig:sd-entable}
\end{figure}

Shower particles at the ground level of the TA site are sampled by the SDs with $3 \, {\rm m^2}$ area. From the lateral distribution of the shower
particles determined from the local densities at SDs \footnote{The measured values are not necessarily 
proportional to number of particles: SDs measure mean energy deposit by shower particles, in
terms of vertical muons.}, an energy estimator, $S_{800}$, the density at 800m from the shower core is evaluated. 
The lateral distribution function of shower particles we used is an AGASA-type formula 
\cite{AGASAlateral},
\begin{eqnarray}
\rho(r) \propto & & \left( \frac{r}{R_M} \right)^{-1.2} \left( 1 + \frac{r}{R_M} \right)^{-(\eta-1.2)}  \nonumber \\
 &  & {} \times  \left[  1 + \left( \frac{r}{1000{\rm m}} \right)^2 \right]^{0.6}
\end{eqnarray}
and the reconstruction procedures are adjusted to fit the TA SD data \cite{ThomsonSD}.
In order to take into account different atmospheric attenuation of inclined showers, we
construct a lookup table for conversion from the observables $(S_{800}, \sec \theta)$ to the energy 
$E_{\rm SD}$ by using a  CORSIKA based Monte Carlo (MC) developed for TA \cite{bibCORSIKA,StokesSD,IvanovSD} (Figure \ref{fig:sd-entable}). 
The SD shower analysis program 
is tuned to reproduce MC-thrown energies from the estimators $(S_{800}, \sec \theta)$ 
of reconstructed showers.


\begin{figure}
\begin{center}
\includegraphics[scale=0.4]{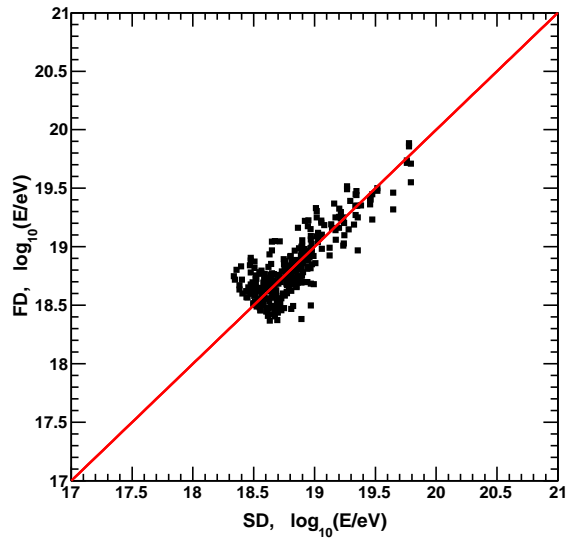}
\end{center}
\caption{Scatter plot of the energies determined from SD and FD reconstructions using the TA hybrid events. Events of all the three combinations (SD-BR/LR/MD) are plotted.}
\label{fig:Esd-Efd}
\end{figure}

However, because of the lack of our knowledge of details of ultra-high energy
hadronic interactions, there are rather large systematic uncertainties in energy determination from
the shower particle measurement at the ground compared to the calorimetric measurement by FDs.
This is a long-standing problem in cosmic ray physics, and has been discussed in comparisons
between AGASA, an SD based experiment, and HiRes, which used FDs.
In order to define a "unified" TA energy scale,
we use {\it hybrid} events, which are detected by both SD and either of the three FD sites. From an 
$E_{\rm SD}^{\rm Hyb} - E_{\rm FD}^{\rm Hyb}$ plot, where 
$E_{\rm SD}^{\rm Hyb}$ and $E_{\rm FD}^{\rm Hyb}$ are energies determined by the SD and FD 
reconstruction of the hybrid events respectively, we determined an overall scale factor 
 $< E_{\rm FD}^{\rm Hyb}/E_{\rm SD}^{\rm Hyb}> = 1/1.27$ to obtain 
the energy of each SD event (Figure \ref{fig:Esd-Efd}). Therefore the energy of an SD event, which is not necessarily a hybrid event,
is given by $E = < E_{\rm FD}^{\rm Hyb}/E_{\rm SD}^{\rm Hyb}> E_{\rm SD}$, where $E_{\rm SD}$ is the energy determined from the SD reconstruction \cite{TsunesadaEnergyScale,IvanovSD}. Note that the hybrid events of
all the three combinations (SD and either of BR/LR/MD) are plotted in Figure \ref{fig:Esd-Efd}. This indicates that the
energy scales of all the three FD stations are in consistent even though the entirely different detectors and calibration
schemes are used.

\section{Energy spectrum}
\begin{figure}
\begin{center}
\includegraphics[scale=0.45]{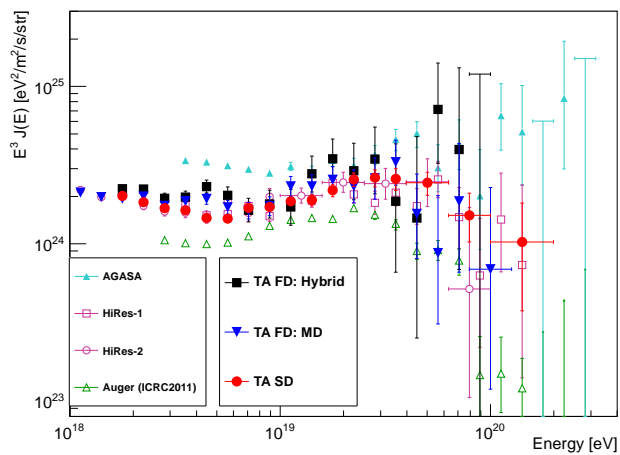}
\end{center}
\caption{TA energy spectra from SD (filled circle), FD mono (MD site, filled triangle), and BR/LR hybrid (filled square).}
\label{fig:spectra}
\end{figure}

In Figure \ref{fig:spectra}, I show three energy spectra above $10^{18}$ eV from the TA data analyses, of the MD monocular mode \cite{DougMD}, the BR/LR {\it hybrid} mode \cite{IkedaHybrid}, and the SD analysis \cite{IvanovSD}.
Here the "BR/LR hybrid" means that we used FD data of BR or LR site, and also the shower particle arrival timing at the position of an SD, which significantly improves the accuracy in determining shower geometries than in the monocular mode \cite{IkedaHybrid}.
For comparison, the AGASA \cite{AGASA}, HiRes \cite{HiRes} and the Pierre Auger Observatory (PAO) \cite{AugerSp} spectra are also
displayed. We stress that all the three TA spectra are in good agreement (note that the SD energy has been
scaled by $1/1.27$ matching to the FD energy scale as described in the previous section). It is also important that the
MD data, from the refurbished HiRes-1 detectors, analyzed with the identical procedures used in HiRes, 
 is fully consistent with the HiRes-1 and HiRes-2 results.

We applied a broken-power-law fit with two bending points to the SD spectrum, which is of the largest exposure in TA,
\begin{eqnarray}
 \frac{{\rm d}I}{{\rm d}E}(E) = A \times \begin{cases} 
  E^k, & (E < E_{\rm a}) \\
  E_{\rm a}^{\ell - k} E^{\ell}, & (E_{\rm a} \le E < E_{\rm c}) \\
  E_{\rm a}^{\ell-k} E_{\rm c}^{\ell -m} E^m, & (E > E_{\rm c})
  \end{cases}
\end{eqnarray}
where $k, \ell, m$ are spectral indices, $E_{\rm a}, E_{\rm c}$ are the bending points, and $A$ is
a normalization constant. The fit parameters are listed in Table \ref{tab:SDspectrum}. The "ankle" structure
is found at $\log_{10} (E_{\rm a}/{\rm eV}) = 18.69$, and the second bending point is 
also identified at $\log_{10} (E_{\rm c}/{\rm eV}) = 19.68$. The number of events observed above
$E_{\rm c}$ is $N_{\rm obs} = 28$, while expected $N_{\rm exp} = 54.9$ in an absence of the second bend. 
The significance of the event deficit above the "cut-off" energy $E_{\rm c}$ is evaluated as
$\sum_{i = 0}^{28} {\rm Poisson}(\mu = 54.9; i) = 4.75 \times 10^{-1} \sim 3.9 \sigma$.

\begin{table}
\begin{center}
\begin{tabular}{ccc}
{\bf $A[{\rm m^{-2} str^{-1} s^{-1} eV^{-1}}] $} & {\bf $\log_{10} (E_{\rm a}/{\rm eV})$}  & {\bf $\log_{10} (E_{\rm c}/{\rm eV})$} \cr \hline
$(2.4 \pm 0.1)  \times 10^{-30} $ &  $18.69 \pm 0.3$ & $19.68 \pm 0.9$ \cr 
  &  & \cr
{\bf $k$} & {\bf $\ell$}  & {\bf $m$} \cr  \hline
$-3.33 \pm 0.4$ & $-2.68 \pm 0.4$ & $-4.2 \pm 0.7$
\end{tabular}
\end{center}
\caption{TA-SD spectral parameters}
\label{tab:SDspectrum}
\end{table}

\section{Mass composition}
The information on the primary nuclear types of cosmic rays can be obtained from the longitudinal
developments of air showers initiated by them. The fluorescence detector is a powerful tool for this purpose, since it
measures shower profiles at different atmospheric depths with many photo-tube "pixels". Here we used BR/LR {\it stereo}
events, which were detected at the two sites, because their shower geometries can be determined more 
accurately than in case of the monocular mode. We used $\Xmax$, the maximum development point of an air shower,
as the mass indicator: $\Xmax$ is larger (deeper in the atmosphere, lower altitudes) for protons or lighter nuclei,
and smaller (higher altitudes) for heavier nuclei as iron. We compared the distribution of $\Xmax$ of observed
showers and those predicted from a shower Monte-Carlo using the CORSIKA code \cite{bibCORSIKA}.
A caveat should be made here: since we observe cosmic rays at the fixed ground levels with the limited field of view,
the distribution of observed $\Xmax$ should be different from a prediction by a shower generator 
like CORSIKA even if the primary composition is fully known, unless applying a sophisticated data selection
\footnote{The PAO uses well-tuned data cuts which give a "true" $\Xmax$ distribution from the
observed data \cite{AugerXmax}.}. We take another approach here. We use a "standard" data selection rule 
for the stereo events. Instead of finding a special selection filter for the data, 
we apply our detector and triggering simulation, 
and also the selection criteria to the CORSIKA showers exactly same as to the data \cite{TamedaXmax1}. The distributions of
$\Xmax$ of such Monte-Carlo  showers with the detector simulation and the cuts can be 
compared with those of the real data.

\begin{figure}
\begin{center}
\begin{tabular}{cc}
\includegraphics[scale=0.2]{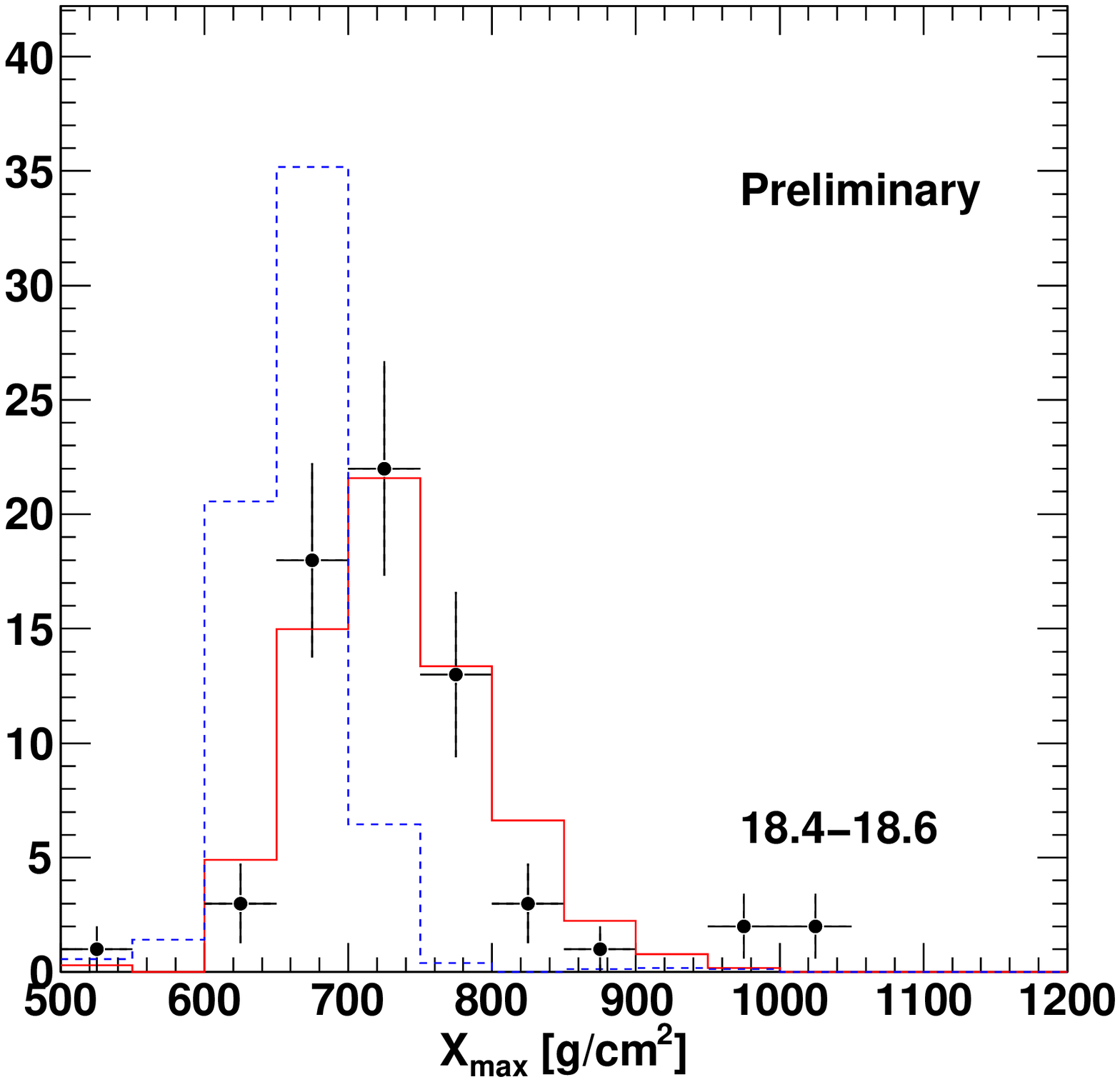} & \includegraphics[scale=0.2]{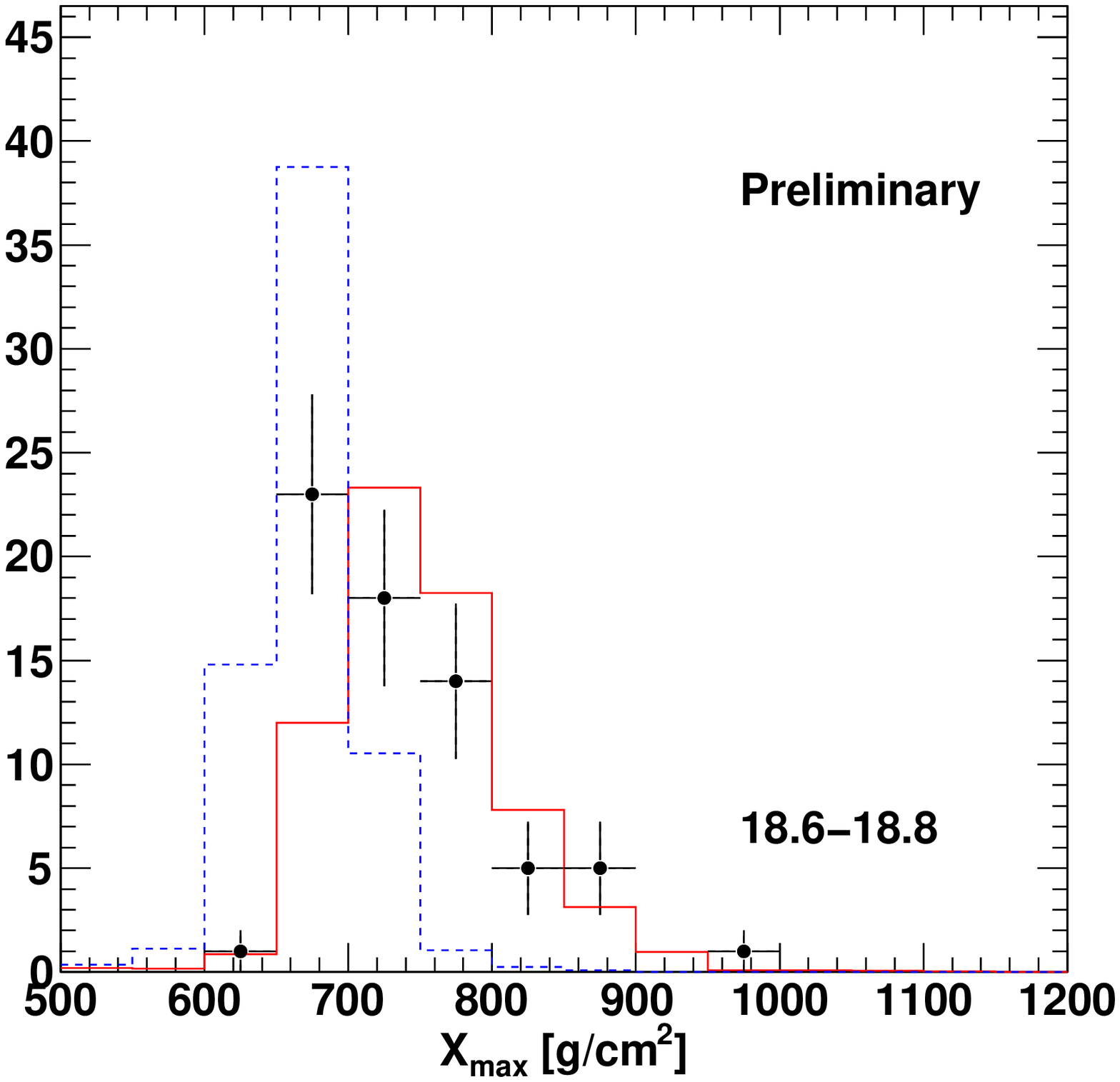} \cr
\includegraphics[scale=0.2]{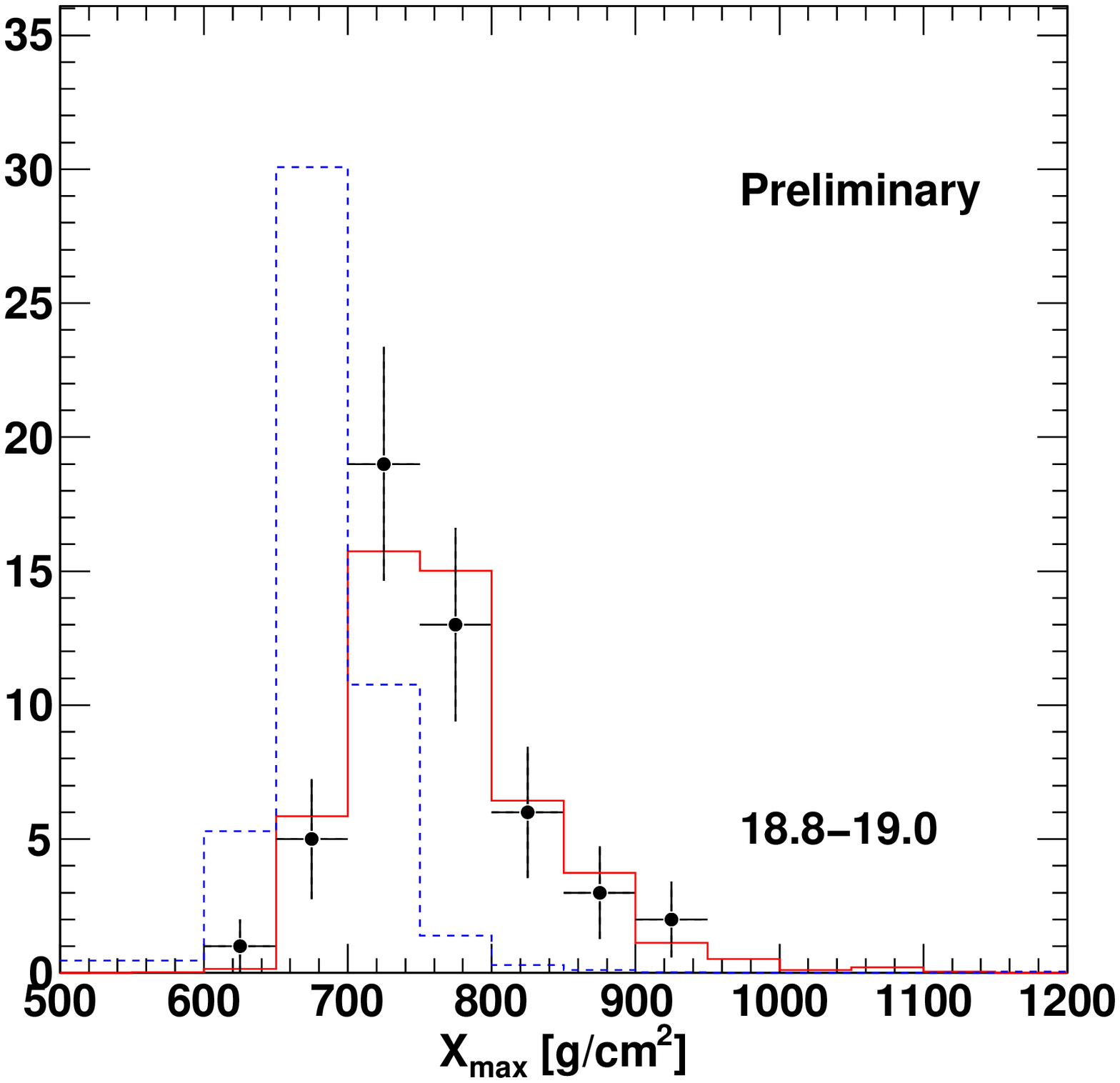} & \includegraphics[scale=0.2]{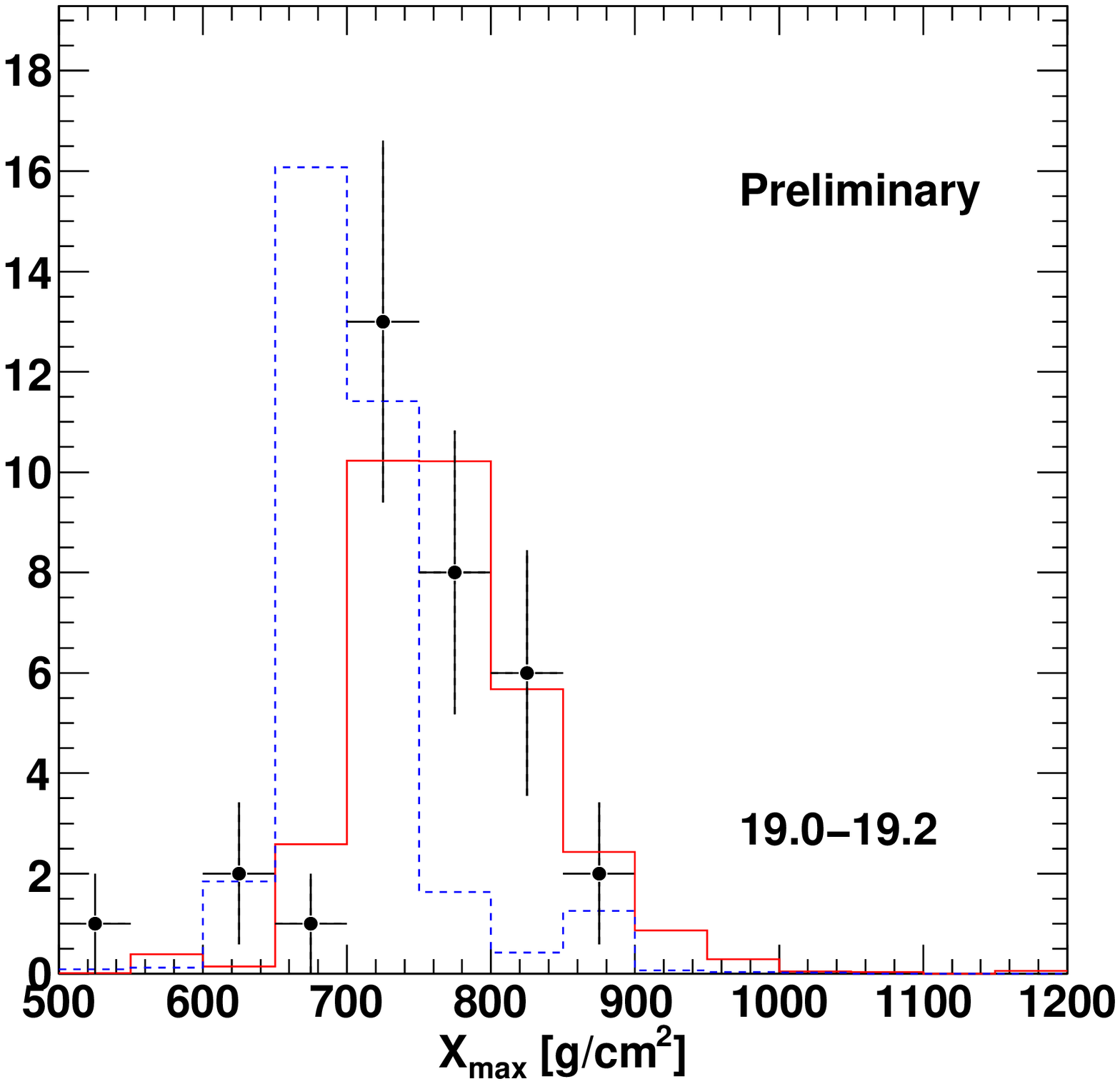} \cr
\includegraphics[scale=0.2]{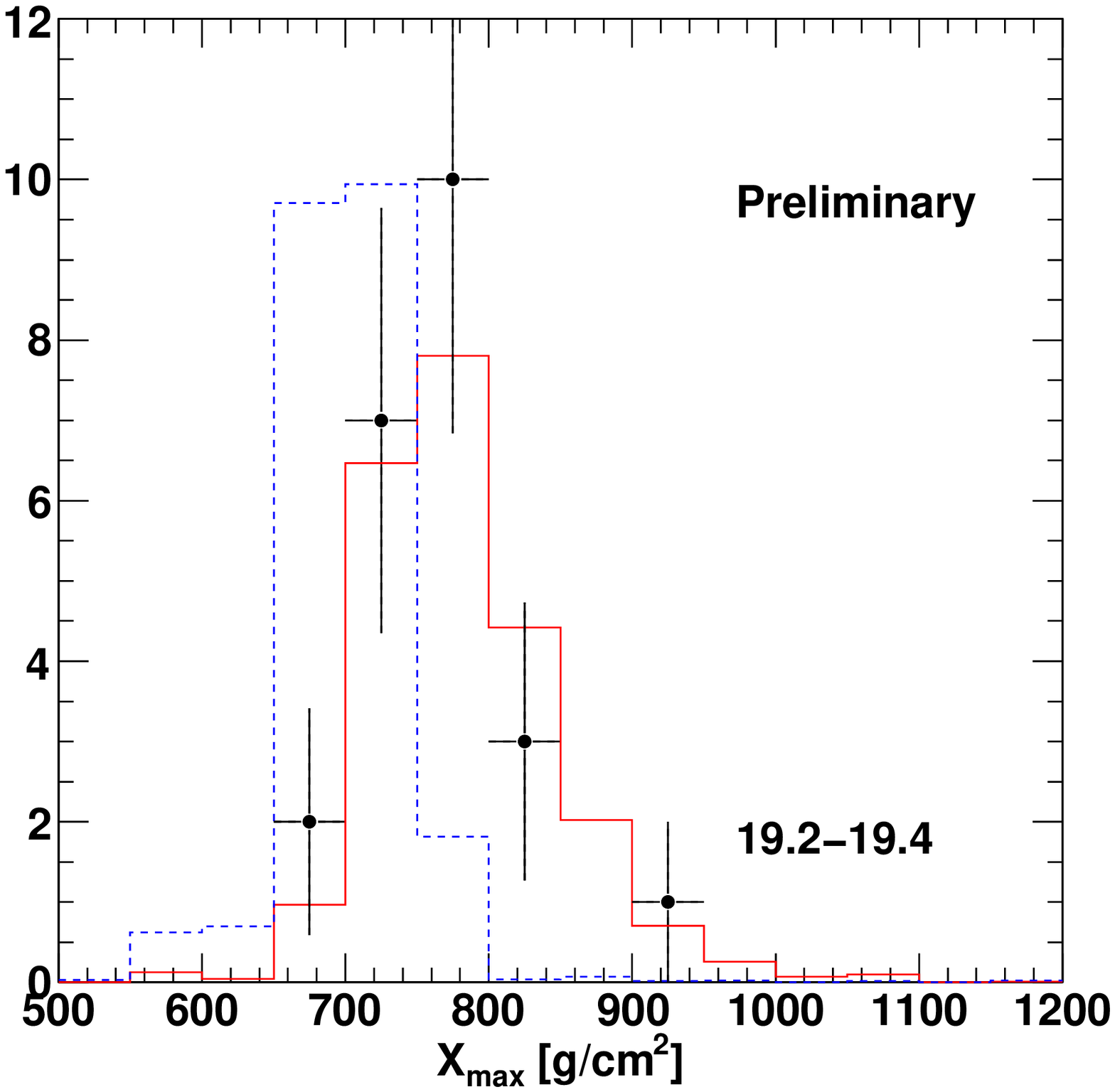} & \includegraphics[scale=0.2]{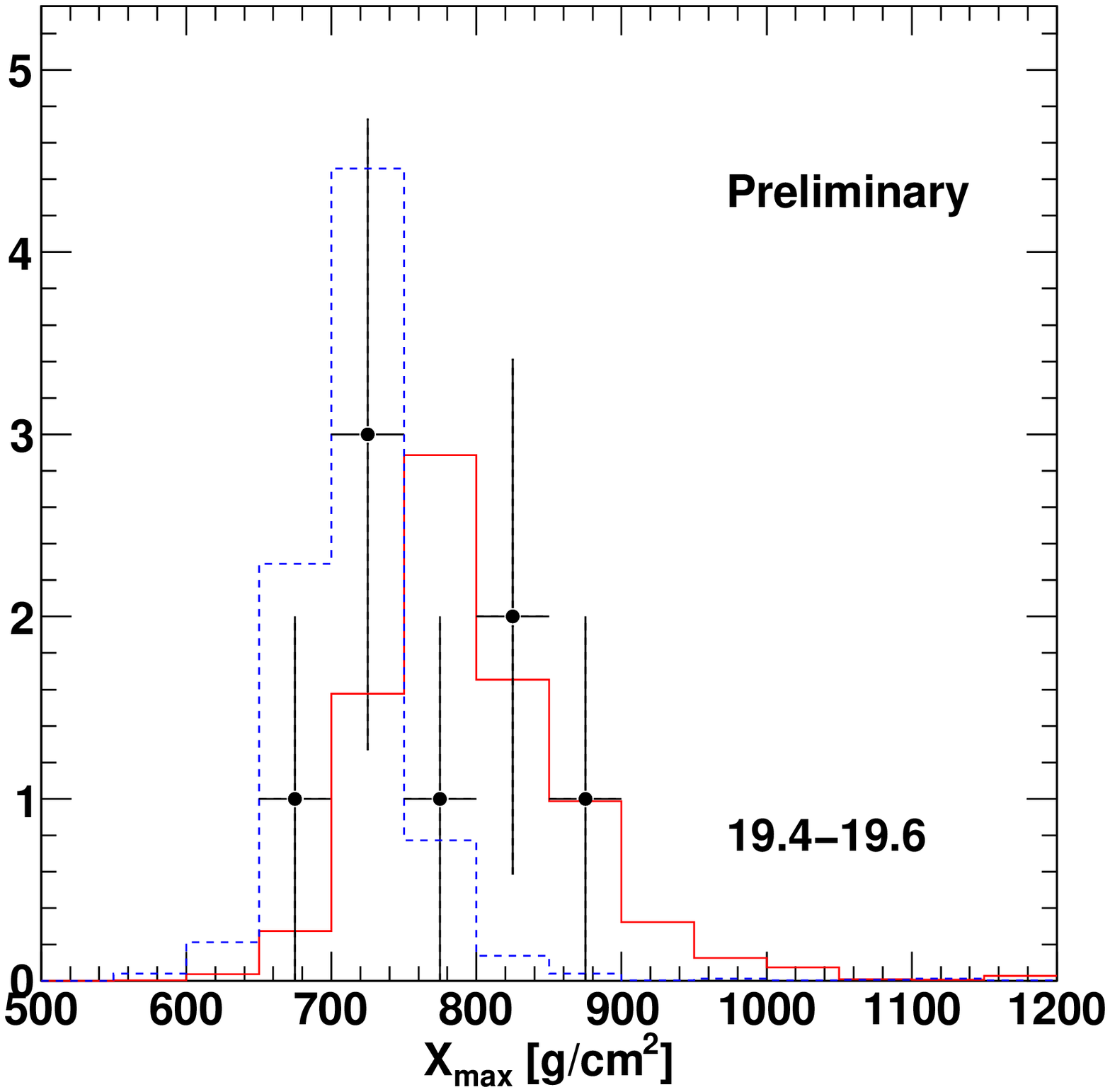} 
\end{tabular}
\end{center}
\caption{$\Xmax$ distributions of the observed showers of the different energy regions (markers), together with the expected ones predicted from our Monte-Carlo studies using CORSIKA showers with the TA-FD detector simulations (histograms: solid for protons and dashed for iron nuclei). The QGSJET-II hadronic interaction model is used.}
\label{fig:XmaxDist}
\end{figure}

The distributions of $\Xmax$ of the observed showers for different energy regions in $E = 10^{18.4 \sim 19.6}$ eV
are shown in Figure \ref{fig:XmaxDist} \cite{TamedaXmax2}. 
The expected distributions for pure-proton and pure-iron primaries are also shown, 
obtained from our Monte-Carlo studies using
the CORSIKA showers with the TA-FD detector simulations and the same data cuts . In order to examine the
compatibilities of the observed $\Xmax$ and the predictions,
KS tests were carried out calculating the "p-values" as functions of energies.
Here we used CORSIKA shower libraries of three different hadronic interaction models, QGSJET-01, QGSJET-II,
and Sibyll. In all the energies and the hadronic interaction assumptions, 
the TA data are compatible with the proton primary predictions. 
In the higher energies ($E > 10^{19.4}$ eV),
the p-values are also high for the iron hypothesis, but the statistics are still limited. 

\begin{figure}
\begin{center}
\begin{tabular}{cc}
\includegraphics[scale=0.44]{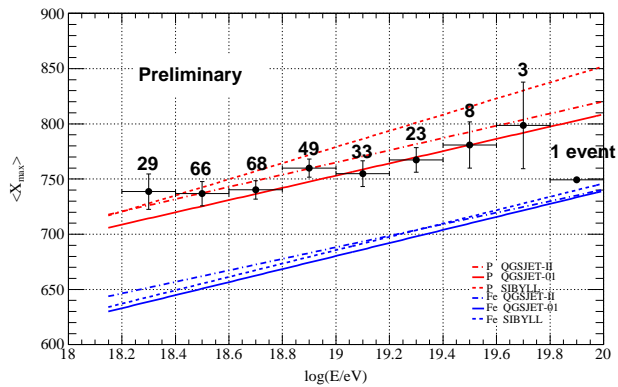} 
\end{tabular}
\end{center}
\caption{$\left< \Xmax \right>$ plot as a function of energy. The proton and iron prediction lines are obtained through the TA-FD detector simulation and the cuts exactly same for the data.}
\label{fig:Xmax}
\end{figure}

I show the averaged $\Xmax$ as a function of energy in Figure \ref{fig:Xmax} \cite{TamedaXmax2}. 
One can see again that the TA data is consistent with the proton-dominated composition of ultra-high energy
cosmic rays. Note that the proton and iron primary prediction lines in Figure \ref{fig:Xmax} are also obtained by applying the TA
detector simulation and the same data cuts for the data, therefore both the data and the prediction lines in this figure
cannot be directly compared with the results from other experiments, like HiRes or PAO.

\section{Anisotropies}
The major difficulty in exploring the origin of cosmic rays is the magnetic deflections of charged particle 
trajectories from their sources to the Earth. However, {\it anisotropies} in the cosmic ray arrival direction distribution 
still give important clues, by finding hot spots of cosmic rays or correlations with possible source populations.
We used the TA-SD events collected from May 2008 to April 2011, of zenith angles smaller than $45^{\circ}$
and angular accuracies $\sim 1.5^{\circ}$. Note that the SD event energies have been renormalized to the FD energy scale ($1/1.27$) as described in the Section \ref{sec:SDanalysis}. The data set contains $854$ events with energies
greater than $10$EeV, and $49, 20$ for $E > 40$EeV, $> 57$EeV, respectively.

\begin{figure}
\begin{center}
\begin{tabular}{cc}
\includegraphics[scale=0.6]{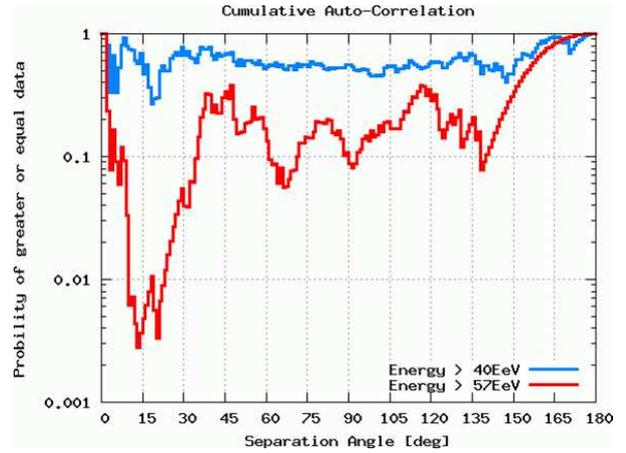} 
\end{tabular}
\end{center}
\caption{Auto-correlation analysis: the vertical axis is probability that the uniform distribution of cosmic rays accidentally gives the same or a greater number of event pairs than the observed number of pairs in a given angular separation.}
\label{fig:autocorr}
\end{figure}

First we examined auto-correlation of the TA events, or event clusterings as reported by the AGASA group \cite{AGASAclustering},
without making any assumptions on cosmic ray sources. 
The number of event pairs in a given angular window $\delta$ is counted, and calculated a probability $P(\delta)$ 
that the uniform
distribution of cosmic rays yields the same or a greater number of pairs by chance than the observed one. The result
is shown in Figure \ref{fig:autocorr} as a function of the window size \cite{TkachevAGN}. There is a rather small $P(\delta)$ region around 
the angular scale $\delta \sim 15^{\circ}$ for $E > 57$ EeV, but not statistically significant to support a deviation
from the isotropy.

\begin{figure}
\begin{center}
\includegraphics[scale=0.3]{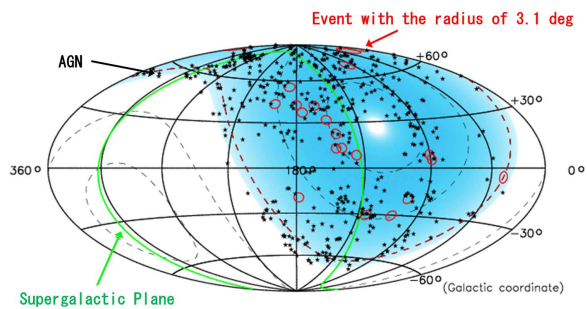}
\end{center}
\caption{20 TA events of $E > 57$ EeV and the AGN in the VCV catalog \cite{VCV}. Galactic coordinates.}
\label{fig:gal-map}
\end{figure}

\begin{figure}
\begin{center}
\includegraphics[scale=0.45]{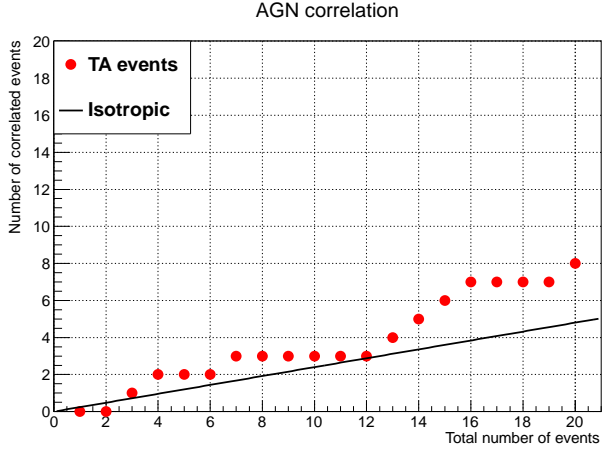}
\end{center}
\caption{The growth of the number of TA events ($E > 57$ EeV) correlating with AGN since 2008.}
\label{fig:agn-corr}
\end{figure}

Second, we searched for cross-correlation of the TA events and active galactic nuclei (AGN), which have been
considered as possible cosmic ray production sites.
The procedures  and parameters used in this analysis  are similar to those
used in the studies by PAO \cite{AugerAGN}. We consider AGNs, QSOs and BL Lacs in the
object catalog (the Veron-Cetty \& Veron (VCV) catalog \cite{VCV})
with redshift smaller than 0.018, and search for correlation with the TA events of $E > 57$ EeV (20 events) within 
an angular separation of $3.1^{\circ}$ (Figure \ref{fig:gal-map}). We assumed that all the objects have the same intrinsic cosmic ray intensities
regardless of class or distance. The number of TA events correlating with the VCV AGN is shown in Figure \ref{fig:agn-corr}, as a function of the total number of events with energies greater than $57$ EeV, in an order of the
detection date \cite{TkachevAGN}. We found 8 correlated events out of 20, while we expect 5 ($25 \%$) chance correlation 
if the source distribution were isotropic in the TA sky. The statistical significance of a deviation
from the isotropy is also small in this analysis, therefore
the AGN hypothesis on the cosmic ray origin is not supported by the present TA data.

\begin{figure}
\begin{center}
\begin{tabular}{cc}
\includegraphics[scale=0.22]{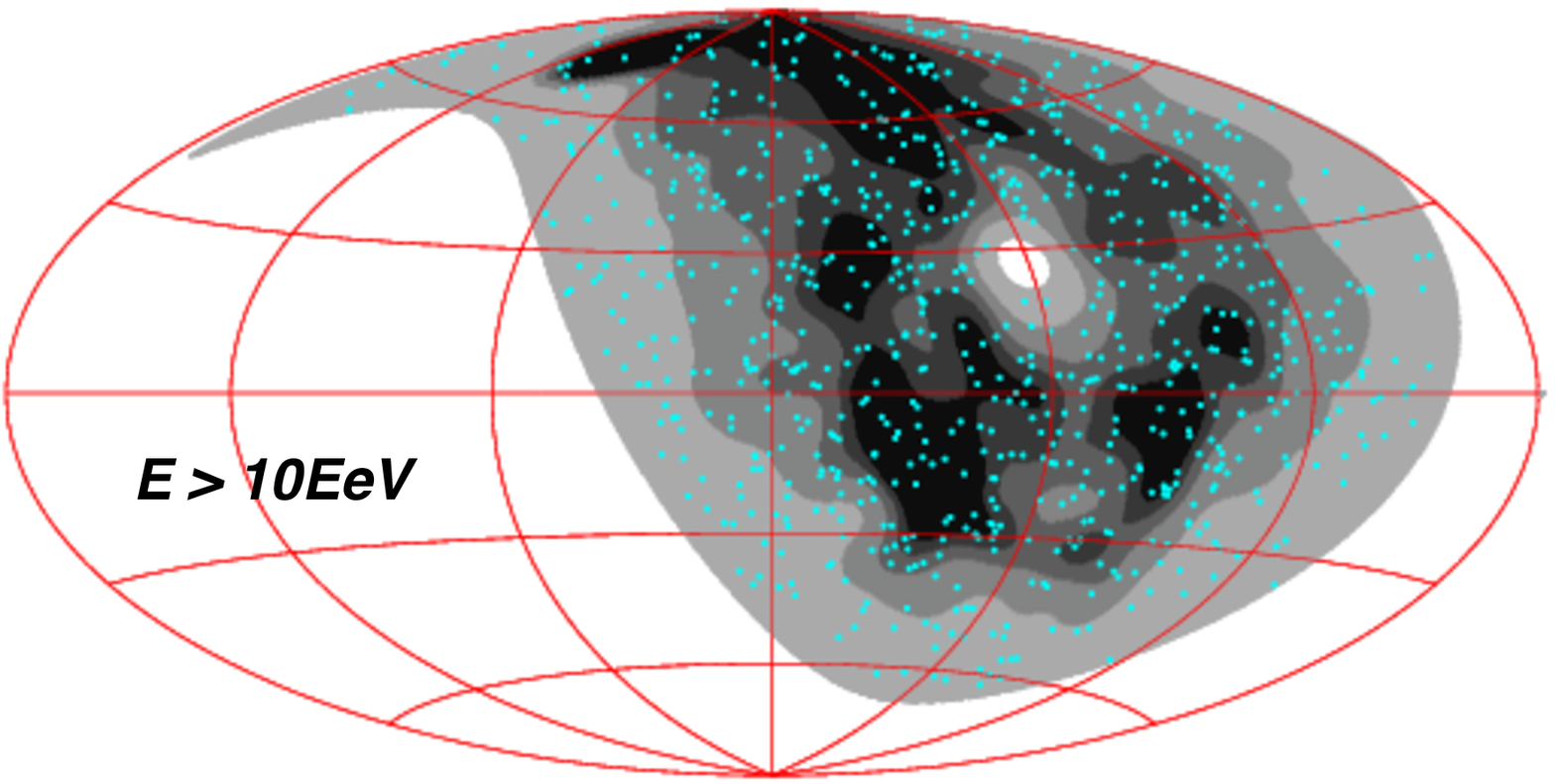} 
& \includegraphics[scale=0.22]{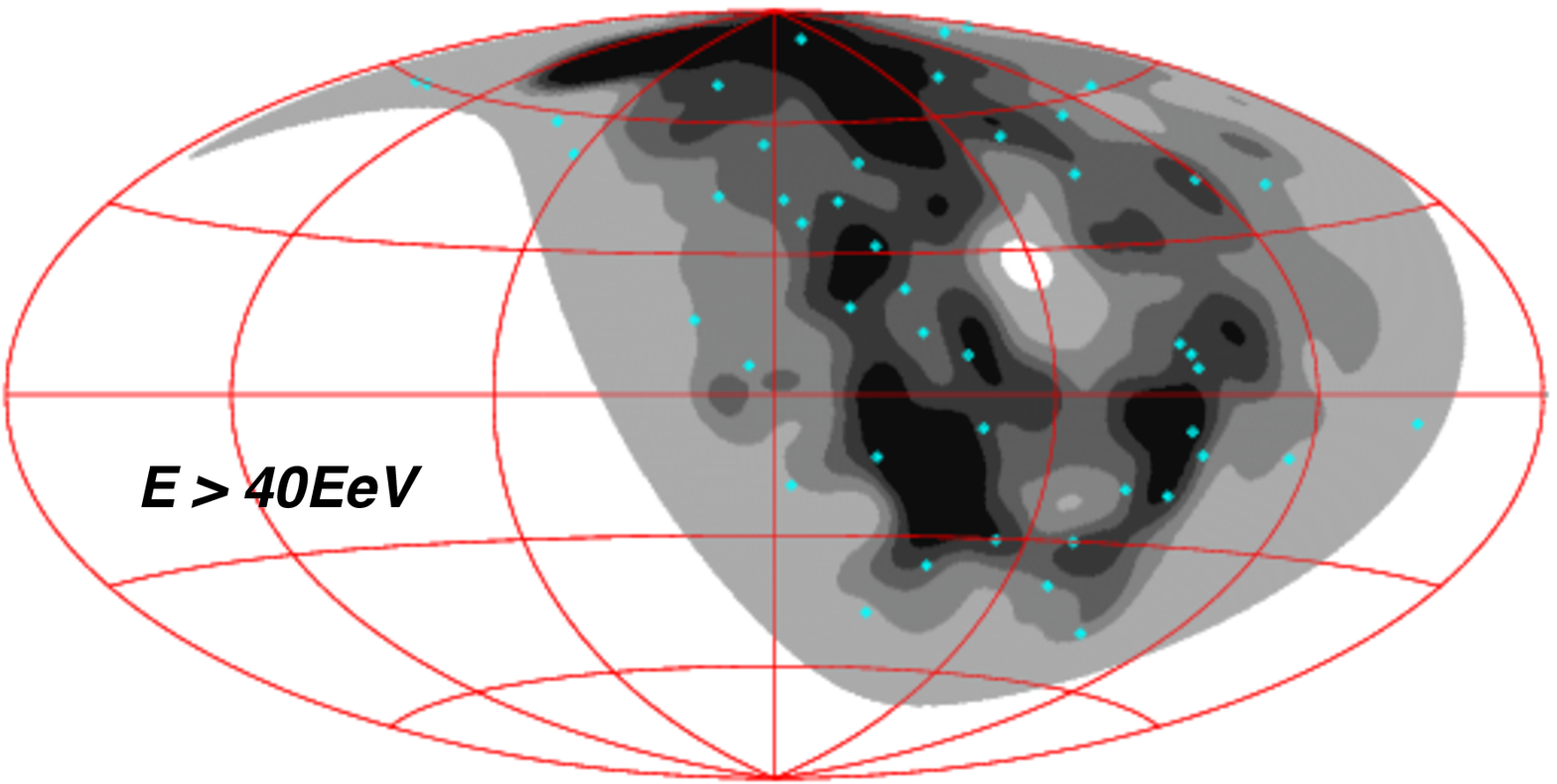} \cr
\includegraphics[scale=0.22]{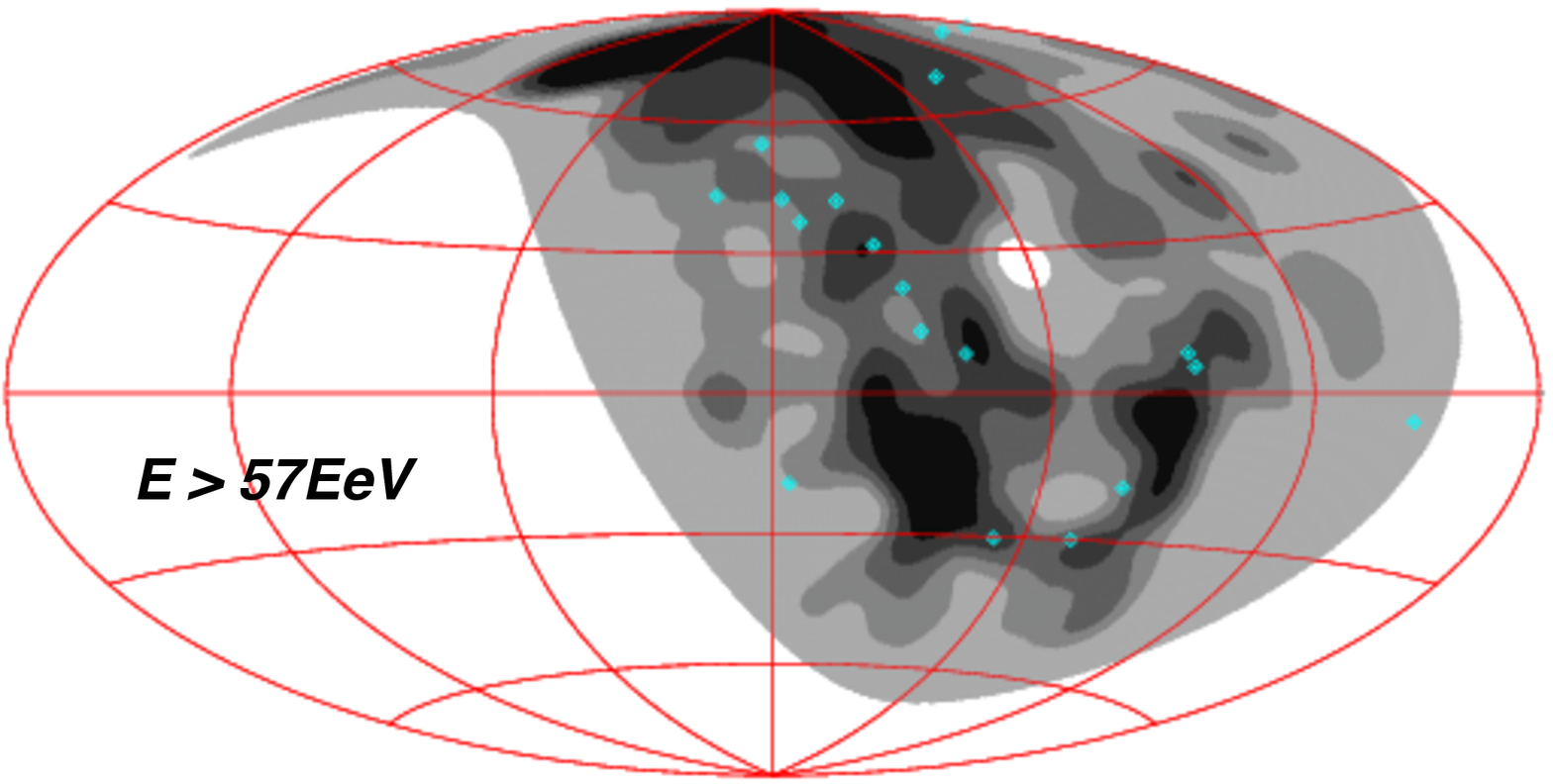} &
\end{tabular}
\end{center}
\caption{The expected cosmic ray density map obtained from the algorithm in \cite{Koers} with the smearing angle $\theta_s = 6^{\circ}$  (contours). The dots are the TA events of the corresponding energies. Galactic coordinates ($\ell = 0$ at the right edge of the figure, increasing $\ell$ to the left, $180^{\circ}$ at the center, and $360^{\circ}$ at the left edge).}
\label{fig:LSSmap}
\end{figure}

\begin{figure}
\begin{center}
\begin{tabular}{cc}
\includegraphics[scale=0.3]{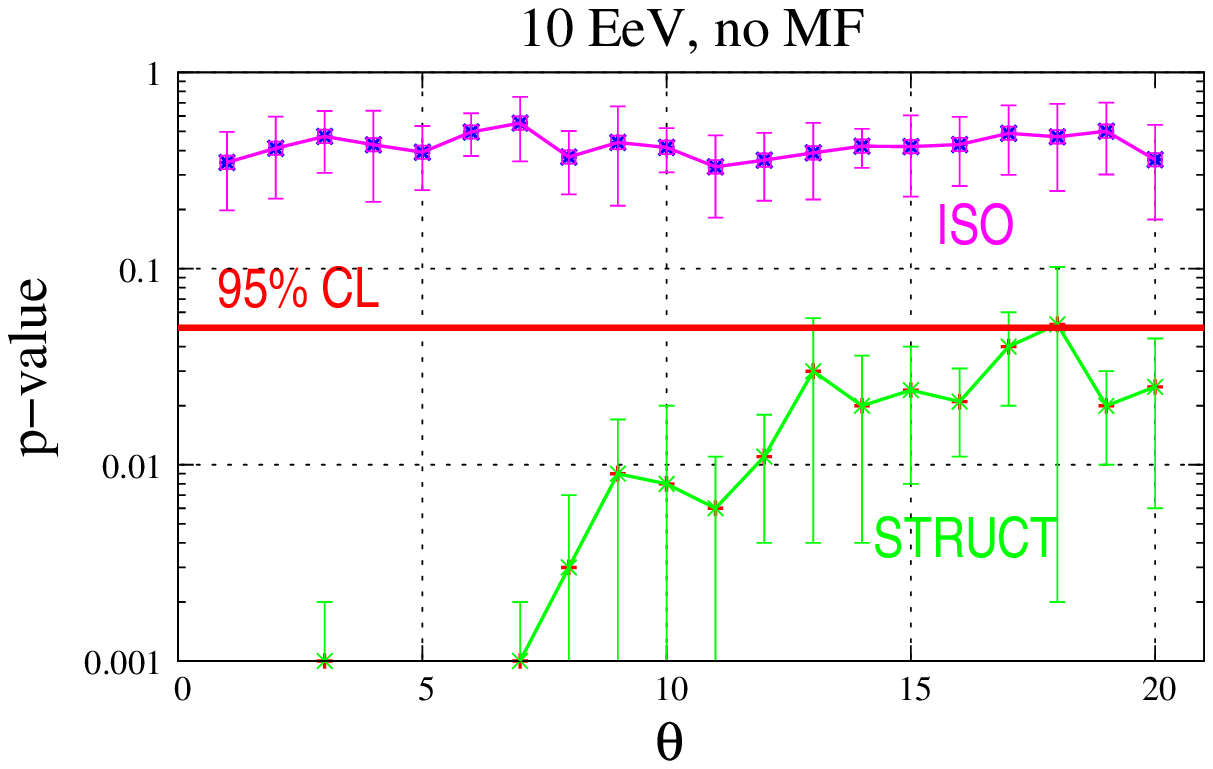}  &
\includegraphics[scale=0.3]{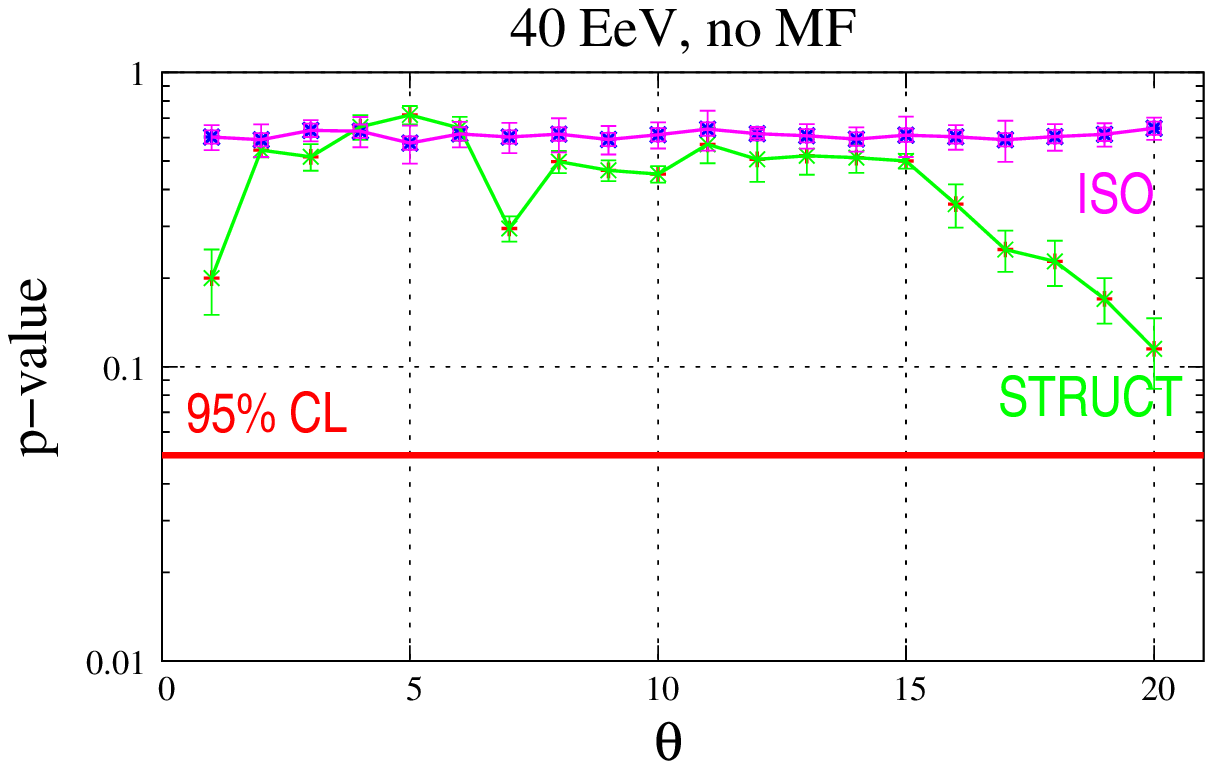} \cr
\includegraphics[scale=0.3]{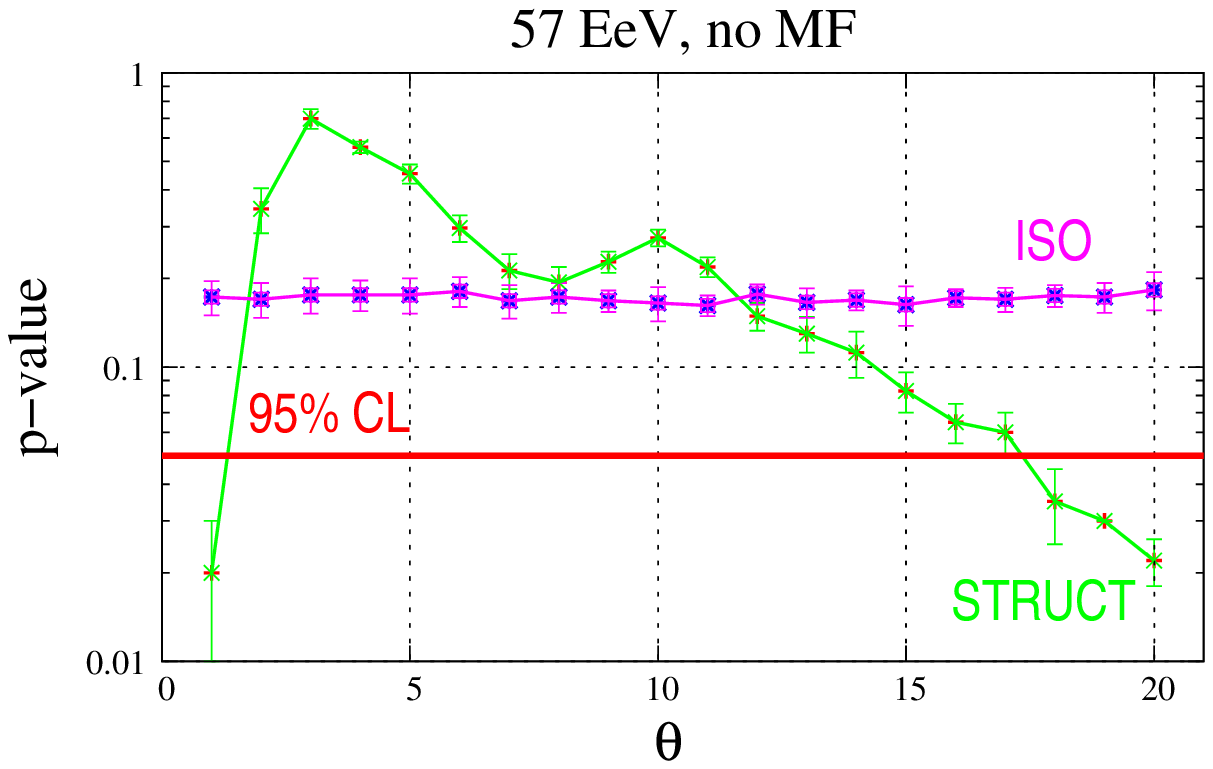} &
\end{tabular}
\end{center}
\caption{The KS test results for the compatibilities between the hypotheses (structure and isotropy) and the TA data. No regular component of the Galactic magnetic field is considered. The horizontal line shows an {\it a priori} chosen confidence level of $95\%$.}
\label{fig:LSS-KS}
\end{figure}

Lastly we investigated a correlation between the TA events and matter distribution in the universe.
We employ the flux sampling method developed in \cite{Koers}, and previously applied to the HiRes data \cite{HiResLSS}. We use 109,408 objects from the 2MASS Galaxy Redshift Catalog (XSCz) \cite{2MASS},
which are located at distances $5  < d < 250$ Mpc from the Earth, as tracers of the local matters in the large-scale
structure (LSS) of the universe. The expected cosmic ray density map is constructed as a superposition of the
contributions of the individual galaxies, taking into account an angular spreading (or {\it smearing}) as a result of the magnetic deflections by the extra-galactic and Galactic irregular fields, flux attenuations due to cosmic ray interactions (e.g. photo-pion productions,
$e^{\pm}$ pair productions),  and the TA exposure. Here the smearing effect is simply introduced as a Gaussian
spreading with the smearing angle $\theta_s$. The uniform component of cosmic rays, as cumulative contribution
of sources beyond $250$ Mpc, is also added.
We assume that all the primary cosmic rays are protons, in accordance
with the TA composition study \cite{TamedaXmax1,TamedaXmax2}. The obtained flux maps for 
$E > 10, > 40$, and $> 57$ EeV, together with the TA events, are shown in Figure \ref{fig:LSSmap} . The compatibilities
of the LSS hypothesis or the isotropy with the TA events were examined with the KS test. The p-values as a function of the smearing angle are shown in Figure \ref{fig:LSS-KS} for the structure hypothesis, as well as for the isotropic distribution \cite{TinyakovNagoya,TinyakovLSS}.
The TA data of $E > 10$ EeV is isotropic, and not compatible with the structure hypothesis. 
The p-values for $E > 40$ and $> 57$ EeV are rather higher, but there are no significant differences between
the structure and isotropy hypotheses. If we consider regular components of the Galactic magnetic field,
including a halo and a disk fields, the p-values of the structure model is mildly increased, while the compatibility
with the isotropy is also maintained. Therefore we conclude that there are no convincing evidences of correlations
with any classes of source candidates in the present TA data.

\section{Further developments}
\begin{figure}
\begin{center}
\includegraphics[scale=0.25]{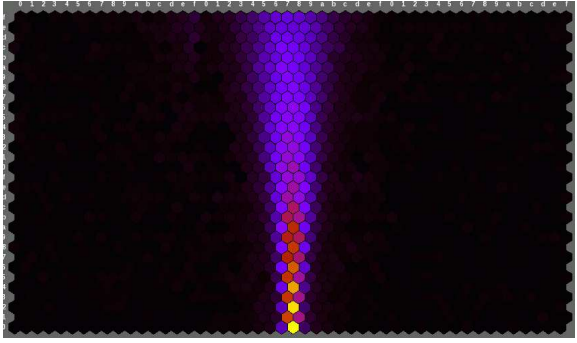}
\end{center}
\caption{An example of the ELS beam shot seen from the TA fluorescence detectors.}
\label{fig:ELSshot}
\end{figure}

\begin{figure}
\begin{center}
\includegraphics[scale=0.43]{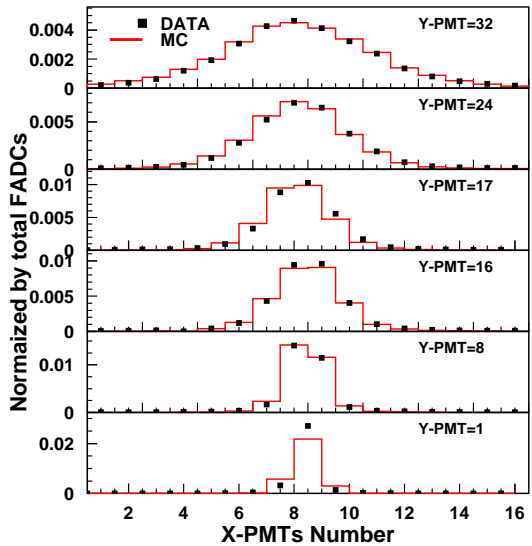}
\end{center}
\caption{Photo-tube signals obtained from an ELS trigger of the fluorescence detectors.}
\label{fig:ELSsignal}
\end{figure}

There are several sources of energy uncertainties in the fluorescence measurements, as atmospheric effects and
fluorescence yields. There have been remarkable progress in laboratory measurements on fluorescence yields,
however, it is not straightforward to apply the results to an actual cosmic ray experiment taking into account all
the atmospheric conditions. An {\it in situ} measurement of fluorescence light using known charged particle energies 
is of great advantage in a cosmic ray experiment, and the Electron Light Source (ELS) of TA is the unique one for this purpose \cite{ShibataELS}. This $40$ MeV electron shooter has been installed at the Black Rock site, 100m away in front of the 
fluorescence detectors, which enables us an end-to-end calibration of our detectors, atmospheric effects, and also fluorescence yields. The ELS has been operational since September 2010 (a beam shot example is shown in 
Figure \ref{fig:ELSshot}). The signals (waveforms) obtained by the photo-tubes which see the electron beams from 
ELS are shown in \ref{fig:ELSsignal}, which are in good agreement with our Monte-Carlo expectations \cite{ShibataELS}. The analysis
of the ELS data is now ongoing, in order to reduce the systematic uncertainty in the FD energies by $\sim 10\%$.

\begin{figure}
\begin{center}
\includegraphics[scale=0.42]{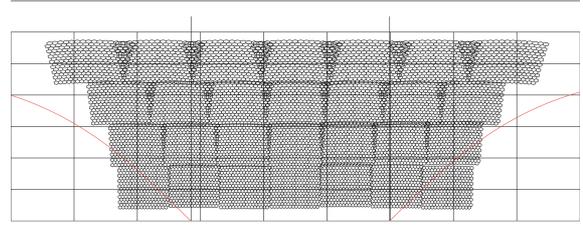}
\end{center}
\caption{TALE-FD field of view}
\label{fig:MD_view}
\end{figure}

\begin{figure}
\begin{center}
\includegraphics[scale=0.3]{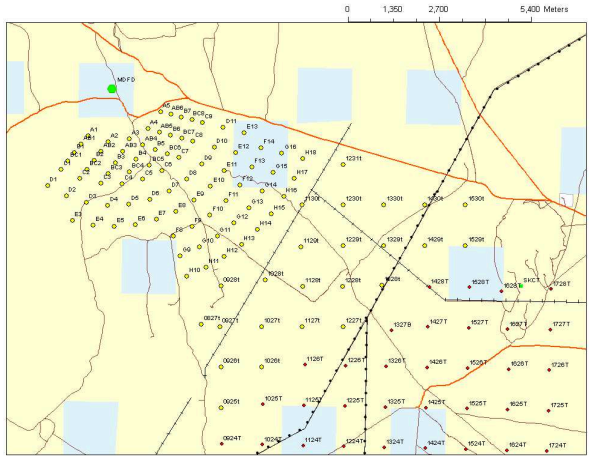}
\end{center}
\caption{Layout of the TALE infill array. The light-colored circles indicated with "A"-H" or 4 digits + "t"  are the TALE counters to be added \cite{TALE}. The other darker circles with 4 digits + "T" represent the existing counters in the
north-west part of the TA SD \cite{NonakaSD}. }
\label{fig:infill}
\end{figure}

I introduce two experimental activities under developments in TA. The one is the TA Low-energy Extension (TALE) \cite{TALE}.
Precise measurements of energy spectrum and mass composition of cosmic rays at energies
$10^{17} \sim 10^{18}$ are quite important by several reasons. In particular, a possible transition of cosmic
rays of Galactic sources to those of extra-galactic origins is of special interest. There is a general agreement
that heavier nuclei are dominant in cosmic rays around the knee region ($E = 10^{15\sim16}$ eV), 
and the average mass increases
with energy up to $\sim 10^{17}$ eV. This is consistent with the supernova-origin scenario of cosmic ray
production and acceleration in our Galaxy.  On the other hand, there are several experimental results
which suggest proton dominant composition above $10^{18}$ eV \cite{GeneralXmax}. Therefore it is expected that there could be a drastic change in mass composition of cosmic rays at $10^{17\sim18}$ eV, which gives a crucial experimental
constraint on astrophysical interpretations of the proposed {\it second knee} or the ankle structures in the
energy spectrum. TALE is designed to lower the minimum detection energy of TA by about an order of magnitude,
with the hybrid detection technique. Reconditioned HiRes-2 detectors as the TALE FD will be installed in the immediate 
vicinity of the MD site i.e. the HiRes-1 detectors. 
The TA-TALE FD covers from $3^{\circ}$ to $59^{\circ}$ in elevation (Figure \ref{fig:MD_view}),
 which enable us to detect lower energy cosmic rays and a bias-free $\Xmax$ measurement.
An infill array is being installed in front of the TA-TALE FDs. A sub-array of 45 counters with a spacing of $400$m
at a distance of $1.5$ km from the TALE-FD gives a $10\%$ efficiency for cosmic rays at $3 \times 10^{16}$ eV.
In order to cover more area, we change the spacing to $600$ in a range of $3 \sim 5$ km from the FD.
For larger distances we use the spacing of the main TA SD array, $1200$ m. The layout of the TALE SD is shown
in Figure \ref{fig:infill}.

\begin{figure}
\begin{center}
\includegraphics[scale=0.27]{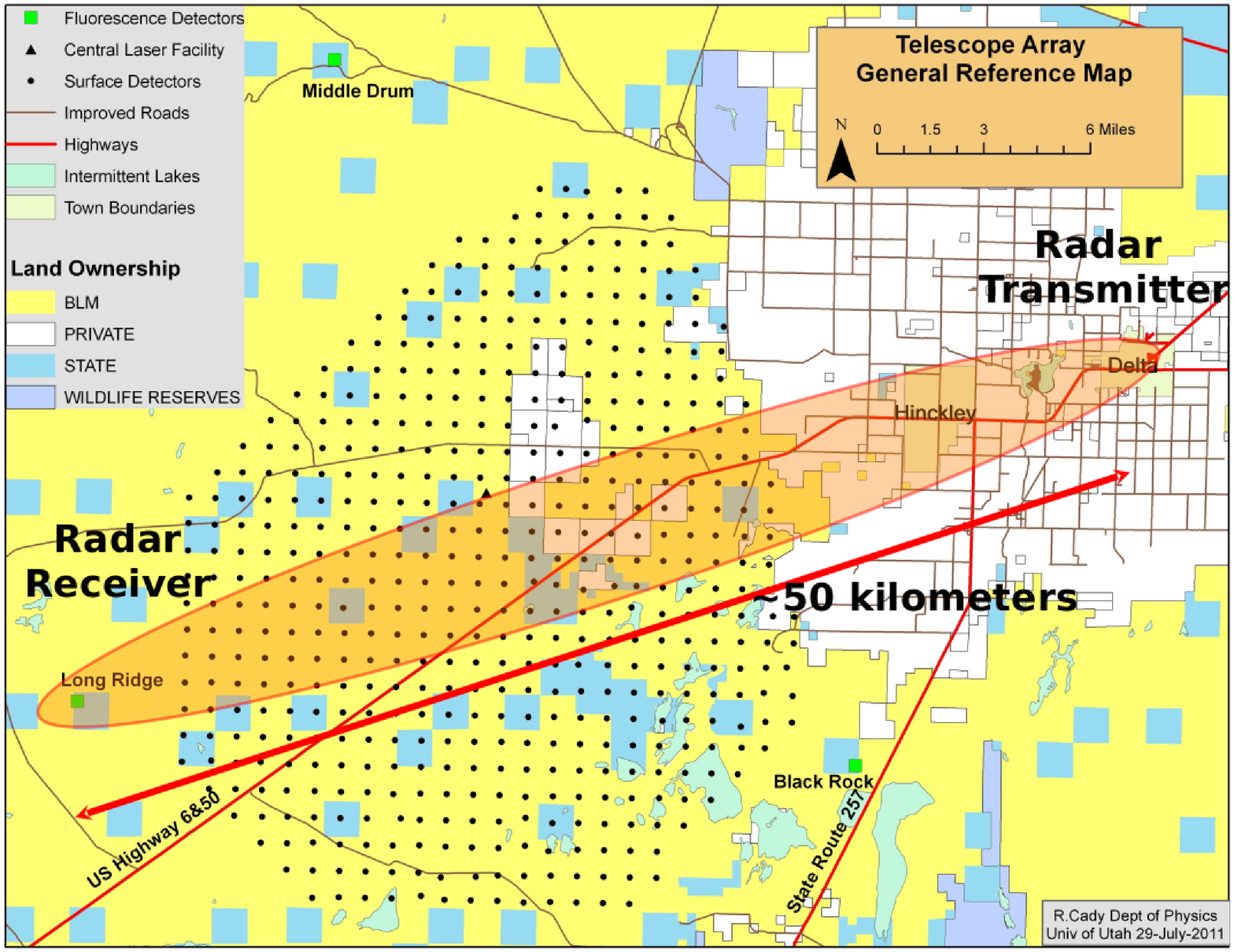}
\end{center}
\caption{TA bistatic radar (TARA): the radio transmitter and the receiver have been installed at a distance of $50$ km.}
\label{fig:TARA}
\end{figure}

\begin{figure}
\begin{center}
\includegraphics[scale=0.27]{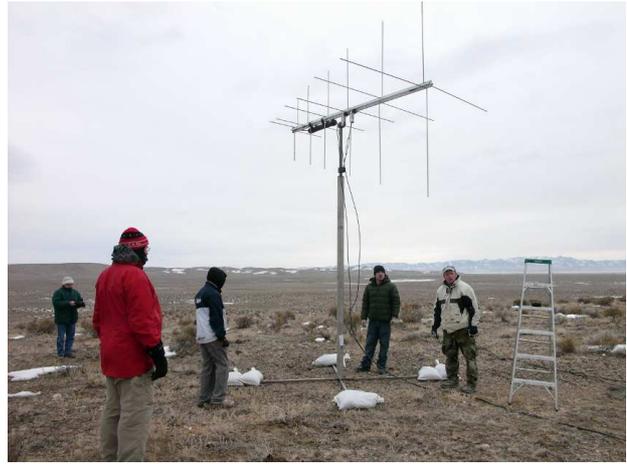}
\end{center}
\caption{TA radar receiver}
\label{fig:RX}
\end{figure}

The another experimental extension in TA is a {\it bistatic radar} detection of ultra-high energy cosmic rays (TARA) \cite{BelzNagoya,TARA}, firstly explored by the MARIACHI project \cite{MARIACHI}. This is a  lower-cost, 24 hour duty 
cycle remote sensing technique for the next generation cosmic ray observatories. 
An important advantage of the bistatic radar technique among the several radio detection methods proposed for
cosmic rays comes from the fact that the scattering cross-section is largest in the  forward direction \cite{RadarCrossSection}:
a radio pulse sent from the transmitter station, and reflected by air shower plasma can be effectively detected
at the receiver station. The signal of the reflected radio pulse has a "chirp"  waveform due to a Doppler-like
effect by the scatterer,  which is the unique feature of the air shower echoes. We have installed a $2$ kW 
analog television transmitter donated from KUTV Channel 2 in Salt Lake City (Figure \ref{fig:TARA}).
This was commissioned in January 2011, operated at a single carrier frequency at 54.1 MHz. The receiver
antennas have been deployed at the TA-LR site, $50$ away from the transmitter station (Figure \ref{fig:TARA}, \ref{fig:RX}).
We also plan to utilize the radar transmitter in conjunction with the TA ELS.
The transmitter will be upgraded at least to $40$ kW by the support from the US-NSF/MRI.

\section{Discussion}

There is a $\sim 20\%$ energy scale difference between TA and PAO, as inferred from the energy spectra. 
It should be noted that there are $22 \sim 23\%$ systematic uncertainties in energy determinations in both experiments, therefore it is possible to say that the two results are consistent within the systematic errors. However, the $20 \%$ difference is not negligible and cannot be overlooked for better understanding of air shower experiment and 
astrophysics of cosmic rays.
It is partly attributed to the different fluorescence yield models employed in the  two groups ($\sim 10\%$), 
but not fully explained only with it.  Some cross-calibration of the detectors will be required to reduce the energy
scale difference, e.g. exchanging the calibration tools or using the TA ELS (this is the author's personal opinion,
and it has not been officially discussed yet between the group headquarters).

Both the TA and PAO $\left< \Xmax \right>$ results show that proton is dominant at energies above $10^{18.2}$ eV.
It can be interpreted that we have not observed an increase of average masse of cosmic rays in the lower energies, 
which is expected from the results of the air shower array experiments targeted at and above the knee region.
This indicates the importance of the lower energy extensions like TA-TALE \cite{TALE},  or AMIGA and HEAT in PAO \cite{AMIGAHEAT}.
It is controversial that there is an apparent difference in the composition studies in TA and PAO in higher energies. 
The TA $\left< \Xmax \right>$ result is consistent with a proton-dominant composition of cosmic rays \cite{TamedaXmax2} (there is one event with a rather smaller
$\left<\Xmax\right>$ at the highest energy bin, but it is of essentially no significance), but an increase of average
cosmic ray mass has been suggested from the PAO data (e.g. \cite{AugerXmax,AugerXmax2,AugerHL}). I propose that we try the same analysis 
procedures: now we are developing a bias-free data cut and determining $\left< \Xmax \right>$ as carried out
in PAO (our preliminary result shows that this brings same amount of shifts in $\left< \Xmax \right>$ to both the
data and Monte-Carlo, therefore the conclusion of proton-dominant composition is not changed at all.). It is possible to employ the TA-style analysis given in \cite{TamedaXmax1,TamedaXmax2} in PAO.
It is also important to revisit the high-energy hadronic interaction models, after the recent experimental
progress by LHC \cite{LHC}.

In 2007, the PAO claimed that the distribution of cosmic ray arrival directions is not isotropic and suggested a possible
correlation with AGN \cite{AugerScience}. However, the statistical significance of the AGN hypothesis in the PAO data 
has been decreased in their updated analysis \cite{AugerAGN2010Nagoya,AugerHL}.
The current TA data shows no statistically significant small or large-scale anisotropies. Unfortunately the interesting
region around Cen A is out of the TA sky. There is no hot spot around the Virgo cluster in the TA data. The absence 
of apparent anisotropy or correlation with known astronomical objects could bring a difficulty in the interpretation
of the spectral features, in particular of the steepening at $\sim 5 \times 10^{19}$ eV, firstly claimed by the HiRes 
experiment \cite{HiResGZK}, and also found in PAO \cite{AugerSpectra} and TA \cite{IvanovSD}. If the primary cosmic rays are protons, and the steepening
is due to cosmic ray energy losses through interaction with the cosmic microwave background photons 
(the Greisen-Zatsepin-Kuzmin effect \cite{Greisen}). If this is the case there could
exist a cosmic ray "horizon" at a certain distance from the Earth, and hence a strong correlation with nearby
objects. Another interpretation is that a heavier component is dominant at the highest energies, therefore the 
steepening we observed is not due to the GZK effect, but to other processes, e.g. acceleration limit at production sites.
In any case, no consistent picture of cosmic ray nature can be drawn from the current observation data, although
we believe that we are just about the answer.
Further efforts in data analysis are required in determining energies and composition, and statistics for
anisotropy studies.

\section{Conclusions}
Telescope Array is the largest cosmic ray observatory in the northern hemisphere, which has been
operational  in full configuration since May 2008. The TA exposure is $\sim 1.7$ AGASA in SD, and
$\sim 1/3$ HiRes in FD. We have developed detector simulators and shower reconstruction programs
for both SD and FD, and the distributions of the observed data are in excellent agreement with expectations
using a previously measured spectrum of cosmic rays. The energy scales of the three FD stations (BR, LR and
MD sites) are also in agreement, even though different detectors and calibration schemes are used.

I presented the major results from the recent progress of the 
TA science, including the energy spectrum,
mass composition, and anisotropy of ultra-high energy cosmic rays. The energy spectrum of cosmic rays
has been derived from analyses of the FD, SD, and the hybrid data. All of the three spectra are in good agreement.
From the spectral fit of the SD spectrum, which is of the largest exposure in TA, we found the ankle structure at
$10^{18.69}$ eV. We also identified a steepening in the spectrum at $10^{19.68}$ eV with a change of the spectral
index from $-2.68$ to $-4.2$. An extended spectrum of cosmic rays at the highest energies is ruled out
by a statistical significance of $3.9 \sigma$. We studied
the distribution of $\Xmax$, at which an air shower has its maximum size, using the FD data. We compared to
the $\Xmax$ distribution of the data and expectations from proton or iron nuclei primaries. The TA data is
consistent with the proton-dominant composition.
Furthermore, we examined the arrival directions of cosmic rays with energies greater than $10$ EeV, and searched
for anisotropies or correlations with known astronomical objects. We concluded that  no statistically significant
anisotropies are found in the present TA data. I introduced a challenge for an end-to-end calibration of the
FDs using the electron accelerator (ELS). I also discussed the further experimental developments in near future, the TA
low energy extension (TALE), and the bistatic radar detection of ultra-high energy cosmic rays with TARA.

It will have been 100 years in 2012 since the discovery of cosmic rays by Victor Hess. The world financial and
political clock is ticking. A critical question to cosmic ray physics is that : "who knows we would discover the origin 
of cosmic rays?" If the TA energy measurement and the composition study are correct, the final unsolved problem
is anisotropy. Our study shows that the statistical power of distinguishing different hypothesis on cosmic ray origins
are still weak, and smaller than $50\%$ in the current TA data \cite{TinyakovNagoya}. 
This will be significantly improved if we have statistics doubled or tripled, therefore we need observations
several more years. We'll also have to discuss the next generation cosmic ray observatory with a huge coverage, 
to explore the astronomy by charged particles.

\section*{Acknowledgement}  
The Telescope Array experiment is supported 
by the Japan Society for the Promotion of Science through
Grants-in-Aid for Scientific Research on Specially Promoted Research (21000002) 
``Extreme Phenomena in the Universe Explored by Highest Energy Cosmic Rays'', 
and the Inter-University Research Program of the Institute for Cosmic Ray 
Research;
by the U.S. National Science Foundation awards PHY-0307098, 
PHY-0601915, PHY-0703893, PHY-0758342, and PHY-0848320 (Utah) and 
PHY-0649681 (Rutgers); 
by the National Research Foundation of Korea 
(2006-0050031, 2007-0056005, 2007-0093860, 2010-0011378, 2010-0028071, R32-10130);
by the Russian Academy of Sciences, RFBR
grants 10-02-01406a and 11-02-01528a (INR),
IISN project No. 4.4509.10 and 
Belgian Science Policy under IUAP VI/11 (ULB).
The foundations of Dr. Ezekiel R. and Edna Wattis Dumke,
Willard L. Eccles and the George S. and Dolores Dore Eccles
all helped with generous donations. 
The State of Utah supported the project through its Economic Development
Board, and the University of Utah through the 
Office of the Vice President for Research. 
The experimental site became available through the cooperation of the 
Utah School and Institutional Trust Lands Administration (SITLA), 
U.S.~Bureau of Land Management and the U.S.~Air Force. 
We also wish to thank the people and the officials of Millard County,
Utah, for their steadfast and warm support. 
We gratefully acknowledge the contributions from the technical staffs of our
home institutions and the University of Utah Center for High Performance Computing
(CHPC). 

\clearpage

\end{document}